\newcommand{\beq}{\begin{equation}}
\newcommand{\eeq}{\end{equation}}
\newcommand{\bea}{\begin{eqnarray}}
\newcommand{\eea}{\end{eqnarray}}

\newcommand{\gsim}{\lower.7ex\hbox{$\;\stackrel{\textstyle>}{\sim}\;$}}
\newcommand{\lsim}{\lower.7ex\hbox{$\;\stackrel{\textstyle<}{\sim}\;$}}

\newcommand{\mrm}{\mathrm}

\documentclass[aps,prev,twocolumn,preprintnumbers,floatfix,nofootinbib]{revtex4-1}
\usepackage{graphicx}
\usepackage{epstopdf}
\usepackage{mathrsfs}
\usepackage{amssymb}
\usepackage{verbatim}

\def\stacksymbols #1#2#3#4{\def\theguybelow{#2}
    \def\vp{\lower#3pt}
    \def\sp{\baselineskip0pt\lineskip#4pt}
    \mathrel{\mathpalette\intermediary#1}}

\def\intermediary#1#2{\vp\vbox{\sp
     \everycr={}\tabskip0pt
     \halign{$\mathsurround0pt#1\hfil##\hfil$\crcr#2\crcr
              \theguybelow\crcr}}}

\def\comment#1{}

\def\u1x{U(1)_X}
\newcommand{\nc}{\newcommand}
\nc{\LL}{L}
\nc{\vv}{\tilde{v}}
\nc{\ccdot}{\!\cdot\!}
\nc{\gsm}{G_{SM}}
\nc{\vfive}{\mathbf{5}\oplus\mathbf{\overline{5}}}
\nc{\vten}{\mathbf{10}\oplus\mathbf{\overline{10}}}
\nc{\zhol}{Z^{\rm hol}}
\nc{\xfb}{\,{\rm fb}}

\begin{document}

%
%

\preprint{LPT--Orsay 10/46}

\title{The kinetic dark-mixing in the light of CoGENT and XENON100}

\author{Yann Mambrini$^{a}$}
\email{Yann.Mambrini@th.u-psud.fr}

\vspace{0.2cm}
\affiliation{
${}^a$ Laboratoire de Physique Th\'eorique \\
Universit\'e Paris-Sud, F-91405 Orsay, France}

\begin{abstract}
Several string or GUT constructions motivate the existence of a $dark$
$U(1)_D$ gauge boson which interacts with the Standard Model only through its
kinetic mixing. We compute the dark matter abundance in such scenario and the
constraints in the light of the recent data from CoGENT, CDMSII and XENON100.
We show in particular that a region with relatively light WIMPS, 
$M_{Z_D}\lsim 40$ GeV and a kinetic mixing $10^{-4} \lsim \delta \lsim 10^{-3} $
is not yet excluded by the last experimental data and seems to give 
promising signals in a near future. We also compute the value of the
 kinetic mixing needed to explain the DAMA/CoGENT/CRESST excesses 
 and find that for $M_{Z_D}\lsim 30 $ GeV, $\delta \sim 10^{-3}$ 
 is sufficient to fit with the data.

\end{abstract}

\maketitle


\maketitle


\setcounter{equation}{0}



\section{Introduction}

Neutral gauge sectors with an additional dark $U(1)_D$ symmetry in addition 
to the Standard Model (SM) hypercharge $U(1)_Y$ and an associated $Z_D$ 
are among the best motivated extensions of the SM, and give the possibility
that a dark matter candidate lies within this new gauge sector of the theory.
Extra gauge symmetries are predicted in most Grand Unified Theories (GUTs)
and appear systematically in string constructions. Larger groups than $SU(5)$
or $SO(10)$, like $E_6$ allows the SM gauge group to be embedded into them.
 Brane--world $U(1)'$s are special compared to GUT $U(1)'$s because there 
 is no reason for the SM particle to be charged 
under them.
For a review of the phenomenology of the extra $U(1)'s$ generated in such scenarios see e.g. \cite{Langacker:2008yv}.
On the other hand, recent anomalies in cosmic rays and direct detection
experiments have motivated the exploration of new gauge interactions
in a putative dark sector \cite{Arkani,Pospelov,Baek,Liu:2009pt}.
The new vector boson $Z_D$ can interact with the SM, even if no 
SM fermions are directly charged under the additional gauge symmetry.
This interaction occurs via mixed kinetic terms between the SM's hypercharge
field strength and the new abelian field strength 
\cite{Holdom,Dienes:1996zr,Martin:1996kn,Rizzo:1998ut,delAguila:1995rb,Dobrescu:2004wz,Cohen:2010kn}. Very recently, a possibility of effective higgs couplings
to the dark sector generated through a triangular loop of $Z$ and/or $Z_D$
has been analyzed in \cite{Cheung} and string scenarios can lead
to naturally light hidden photons \cite{Javier}.  
Other important consequences and 
clear dark matter signatures in satellite telescopes are
studied in \cite{Mambrini:2009ad,Dudasline,Higgsspace}

Our objective is triple:  to know if, taking into account the last data and 
analysis from CDMSII, CoGENT and XENON100, there is still part of the
 parameter space allowed by all constraints, especially WMAP and the 
 electroweak precision tests.
Secondly, considering CoGENT, CDMSII or DAMA as $signal$ events, we 
 compute the kinetic mixing and $Z_D$ mass required to fit the excesses.
   Finally, if we consider that the last XENON100 results exclude
CoGENT and DAMA excesses, we project the future sensitivity needed to explore
the remaining part of the parameter space.   
The paper is organized as follows: after an introduction to the model and 
its motivations, we look in details the cosmological and accelerator 
constraints we should apply for our study.
We then look at the parameter space already reached out by XENON100 and 
CDMS-Si
and fit the last data released by CoGENT, DAMA and CRESST. 
We conclude with some prospects for the XENON100 experiment.

\section{The model}


The matter content of any $dark$ $U(1)_D$ extension of the SM can be decomposed
into three families of particles:

\begin{itemize}
\item{The $Visible$ $sector$ is made of particles which are charged under the SM
gauge group $SU(3)\times SU(2)\times U(1)_Y$ but not charged under $U(1)_D$
(hence the $dark$ denomination for this gauge group)}
\item{the $Dark$ $sector$ is composed by the particles charged under
$U(1)_D$ but neutral with respect of the SM gauge symmetries. The dark matter
($\psi_0$) candidate is the lightest particle of the $dark$ $sector$}
\item{The $Hybrid$ $sector$ contains states with SM $and$ $U(1)_D$ quantum numbers. These states are fundamental because they act as a portal between
the two previous sector through the kinetic mixing they induce at loop
order.} 
\end{itemize}

\noindent
From these considerations, it is easy to build the effective lagrangian
generated at one loop :

\begin{eqnarray}
{\cal L}&=&{\cal L}_{\mrm{SM}}
-\frac{1}{4} \tilde B_{\mu \nu} \tilde B^{\mu \nu}
-\frac{1}{4} \tilde X_{\mu \nu} \tilde X^{\mu \nu}
-\frac{\delta}{2} \tilde B_{\mu \nu} \tilde X^{\mu \nu}
\nonumber
\\
&+&i\sum_i \psi_i \gamma^\mu D_\mu \psi_i
+i\sum_j \Psi_j \gamma^\mu D_\mu \Psi_j
\label{Kinetic}
\end{eqnarray}

\noindent
$B_{\mu}$ being the gauge field for the hypercharge, 
$X_{\mu}$ the gauge field of $U(1)_D$ and
$\psi_i$ the particles from the hidden sector, $\Psi_j$ the particles
 from the hybrid sector, 
$D_{\mu}  =\partial_\mu -i (q_Y \tilde g_Y \tilde B_{\mu} + q_D \tilde g_D
 \tilde X_{\mu} + g T^a W^a_{\mu})$, $T^a$ being the $SU(2)$ generators, and 

\beq
\delta= \frac{\tilde g_Y \tilde g_D}{16 \pi^2}\sum_j q_Y^j q_D^j 
\log \left( \frac{m_j^2}{M_j^2} \right)
\eeq

\noindent
with $m_j$ and $M_j$ being hybrid mass states \cite{Baumgart:2009tn} .

Notice that the sum is on all the hybrid states, as they are the only ones which can contribute to the $Y_{\mu}X_{\mu}$ propagator.
After diagonalization of the current eigenstates that makes the gauge kinetic
terms of Eq.\ref{Kinetic} diagonal and canonical, 
we can write after the $SU(2)_L\times U(1)_Y$ breaking\footnote{Our notation
for the gauge fields are 
($\tilde B^\mu,\tilde X^\mu$) before the diagonalization, 
($B^\mu, X^\mu$) after diagonalization and 
($Z^\mu,Z_D^\mu$) after the electroweak breaking.} :

\begin{eqnarray}
A_{\mu} &=& \sin \theta_W W_{\mu}^3 + \cos \theta_W B_{\mu}
\\
Z_{\mu} &=& \cos \phi ( \cos \theta_W W_{\mu}^3 - \sin \theta_W B_{\mu})
- \sin \phi  X_\mu
\nonumber
\\
(Z_D)_{\mu}&=&\sin \phi (\cos \theta_W W_\mu^3 - \sin \theta_W B_\mu)
+ \cos \phi  X_\mu
\nonumber
\end{eqnarray}

\noindent
with, at the first order in $\delta$:

\begin{eqnarray}
\cos \phi &=& \frac{\alpha}{\sqrt{\alpha^2 + 4 \delta^2 \sin^2 \theta_W}}
~~
\sin \phi = \frac{2 \delta \sin \theta_W}{\sqrt{\alpha^2 + 4 \delta^2 \sin^2 \theta_W}}
\nonumber
\\
\alpha &=& 1- M^2_{Z_D}/M^2_Z - \delta^2 \sin^2 \theta_W
\nonumber
\\
&\pm& \sqrt{1-M^2_{Z_D}/M^2_Z + 4 \delta^2 \sin^2 \theta_W}
\end{eqnarray}

\noindent
and + (-) sign if $M_{Z_D}< (>)M_Z$.
The kinetic mixing parameter $\delta$ generates an effective coupling of 
SM states $\psi_{\mrm{SM}}$ to $Z_D$, and a coupling of $\psi_0$ to 
the SM $Z$ boson which
induces an interaction on nucleons.
Developing the covariant derivative on SM and $\psi_0$ fermions state,
we computed the effective $\psi_{\mrm{SM}}\psi_{\mrm{SM}}Z_D$  and $\psi_0\psi_0Z$ couplings at first order in $\delta$. One can find other implications of such construction in
\cite{Baumgart:2009tn, Pospelov:2007mp, Pospelov:2008zw}

\begin{figure}
    \begin{center}
    \includegraphics[width=1.5in]{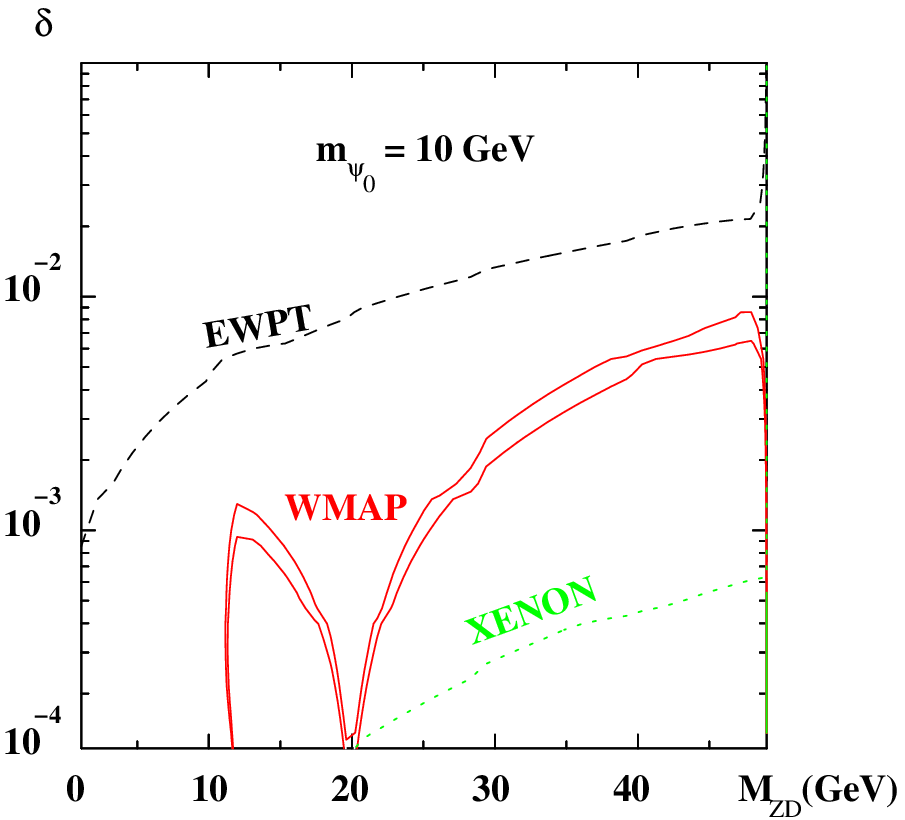}
    \hspace{0.3cm}
    \includegraphics[width=1.5in]{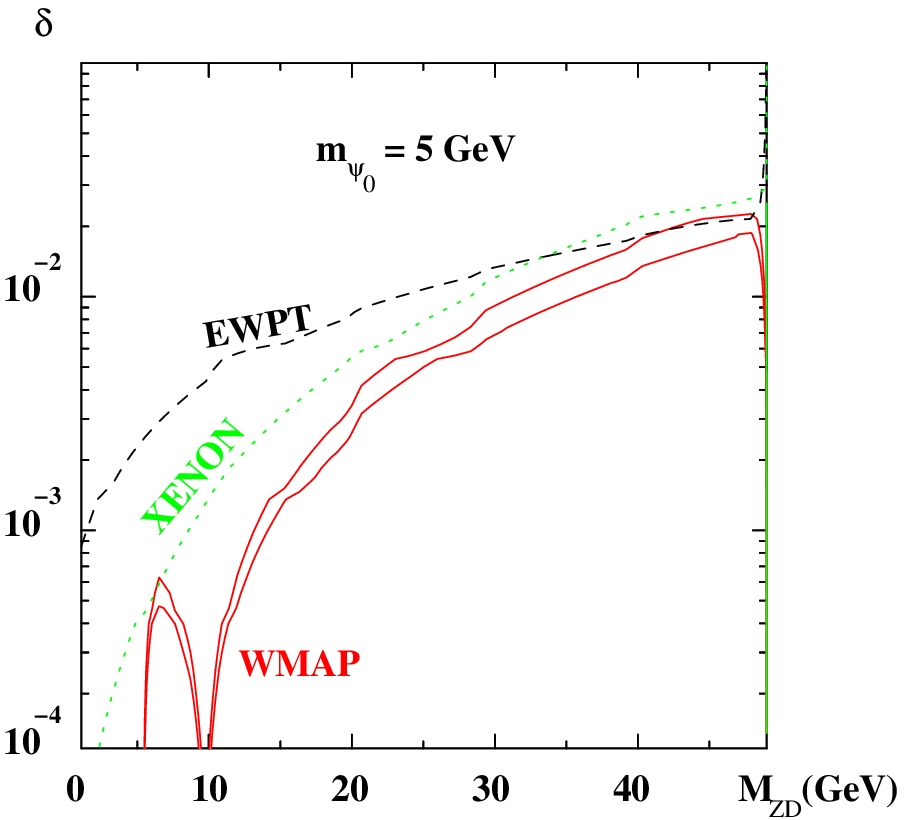}
    
          \caption{{\footnotesize
Two examples of allowed parameter space for $m_{\psi_0}=10$ GeV (left)
and $m_{\psi_0}=5$ GeV (right). The points between the full-red region
respect the 5$\sigma$ WMAP constraint, the points below the dashed-black
line do not exceed accelerator data on precision tests, and the points
above the dotted-green line are excluded by XENON100 data. 
}}
\label{fig:Allowed}
\end{center}
\end{figure}

\section{The constraints}

\subsection{The cosmological constraint}

The abundance of a thermal relic dark matter candidate $\psi_0$ is 
controlled by its annihilation
cross section into SM particles mediated by the exchange of a $Z_D$ gauge
boson through $s-$channel or $t-$channel $Z_DZ_D$ final state 
(see \cite{Mambrini:2009ad,Dudasline} 
for a detailed study of the relic abundance constraints).
 We modified the micrOMEGAs2.2.CPC code\footnote{The author wants to thank 
 particularly G. Belanger and S. Pukhov for their help to address this issue.}
  \cite{Micromegas} in order to calculate 
 the relic abundance of $\psi_0$. We show in Fig.\ref{fig:Allowed} 
 the points that fulfill the WMAP $5\sigma$ bound \cite{WMAP} 
 on $\mathrm{\Omega_{DM}}$ for $m_{\psi_0}= 10$ GeV (left) and 5 GeV (right)
 in the ($M_{Z_D},\delta$) plane.
 One can clearly sees the $Z_D-$pole region when 
 $M_{Z_D} \sim m_{\psi_0}$.
One important point is that for a given $M_{Z_D}$ and $m_{\psi_0}$, 
there exists a unique solution $\delta$ (up to the very small uncertainties at 5$\sigma$)
fulfilling WMAP constraints : from 3 parameters
($m_{\psi_0}, M_{Z_D}, \delta$), the WMAP constraints reduce it to two 
($M_{Z_D}, \delta$).

\subsection{The electroweak precision constraints}

Concerning the electroweak symmetry breaking, the mixing between 
$\tilde X_{\mu}$ and $\tilde B_{\mu}$ generates new contributions 
to precision electroweak observables.
However, none of the particle of the SM has any $U(1)_D$ charges:
the $U(1)_D$ can be considered has a $lepto-hadrophobic$ $Z_D$. 
Other authors in \cite{Feldman:2007wj,Liu:2009pt,Chang:2006fp}  or
\cite{Cassel:2009pu} have looked at hidden-valley like models 
or milli--charged dark matter but concentrating their study to relatively
heavy $Z_D$ and large mixing angle.
The authors of \cite{Kumar:2006gm} 
have computed the observables from effective Peskin--Takeuchi 
parameters\cite{EWPT}, and found

\begin{eqnarray}
\Delta m_W &=& (17 \mrm{MeV}) ~\zeta 
\nonumber
\\
\Delta \Gamma_{l+l-} &=& -(8 \mrm{keV})~ \zeta
\nonumber
\\
\Delta \sin^2 \theta_W^{eff} &=& -(0.00033) ~\zeta
\label{Eq:EWPT}
\end{eqnarray}

\noindent
where

\beq
\zeta \equiv 
\left( \frac{\delta}{0.1} \right)^2
\left( \frac{250 \mrm{GeV}}{m_X} \right)^2
\eeq

\noindent
Different electroweak measurements from LEP give $|\zeta| \lsim 1$.
These constraints are represented by the black line in Fig.\ref{fig:Allowed}. 
A new analysis was made more recently in \cite{Erler:2009ut}
 and \cite{delAguila:2010mx} but they confirmed that in models 
 with extra $U(1)'s$ 
 which does not couple at tree-level with SM particles (like a leptophobic
 $Z_D$  for instance) or with the higgses, the constraints on 
 the mass of the gauge boson are very weak. The only case where one 
 can put some strongest  constraints is if a non-trivial higgs sector acts 
 as a portal between 
 the Dark sector and the SM one at tree level 
 (like in Supersymmetry for instance).
 
 However, for the mass range of interest in this work 
 ($m_{Z_D}\sim 10$ GeV) we needed to look at the search of 
 production/decay of hidden bosons at low energy $e+e-$ colliders
 \cite{EWPT}.
 Indeed, over a large range of parameters, the cross sections for the 
 production of dark--sector particles scale as
 
 \beq
 \sigma \sim \frac{g g_D \delta^2}{16 \pi^2 E^2_{\mrm{cm}}}
 \eeq

\noindent
where $E_{cm}$ is the center-of-mass energy of the collider.
The search sensitivity of a given $e^+e^-$ machine above mass threshold
scales as the ratio of integrated luminosity over squared center-of-mass
energy, ${\cal L}_\mrm{int}/E_{\mrm{cm}}^2$. LEP and Tevatron are much
less sensitive to direct production of low mass dark sectors than the 
B-factories. 

While writing the present article, the authors of 
\cite{Hook} have published an extensive
model independent analysis in the energy range of interest in our study. 
They bounded the kinetic mixing by $\delta \lsim 0.03$ for 
$10 ~ \mrm{GeV}< M_{Z_D} < 200 ~ \mrm{GeV}$ which is in complete agreement
with the constraints given by Eq.\ref{Eq:EWPT} plotted in 
Fig.\ref{fig:Allowed}. Their strongest upper limit  
($\delta \lsim 0.03$)
come from a $wide$ $Z_D$ whose dark decay is maximized. For 
$M_{Z_D}\lsim 10$ GeV, they also computed a model dependent exclusion
 from BaBar searches and obtained $\delta \lsim 3 \times 10^{-3}$ 
 which is also in agreement with Eqs.\ref{Eq:EWPT}.

\subsection{The XENON/ CDMS constraints}

In recent months, there have been new data releases from many experiments
that have engendered a great deal of excitement (see section 
\ref{section:Signals} for a discussion and references).
The XENON100 collaboration has recently released new dark matter limits
\cite{XENON100}, placing particular emphasis on their impact on
searches known to be sensitive to light-mass ($\sim 10$ GeV) WIMPs.
The existing bounds set by the XENON10 \cite{XENON10}, and the 
recognition that the effect of channeling in NaI(Tl) crystal is less
important than previously assumed \cite{CDMS1} can be combined by
the full data set released recently by CDMS \cite{CDMS2} to obtain
tighter bounds to the elastic cross section.
In our work, we will use the analysis made by the authors of
\cite{Kopp:2009qt},
whereas a similar analysis can be found in \cite{Chang:2010yk}.
In practice, the differences between CDMS and XENON constraints appear 
when $ m_{\psi_0} \lsim 10 \mathrm{GeV}$, where CDMS-Si is more sensitive 
than XENON100 [up to the renormalization used for the calculation of
the XENON100 efficiency discussed in the section \ref{section:Signals}].
We show in Fig.\ref{fig:Allowed} two examples of points, one
excluded (left) and the other one allowed (right) by XENON100.

\section{Results}

\subsection{Combining all the constraints}

\begin{figure}
    \begin{center}
    \includegraphics[width=2.7in]{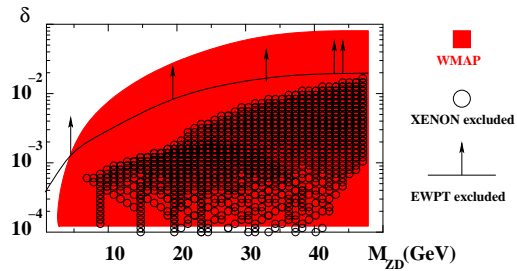}
          \caption{{\footnotesize
Constraints coming from WMAP (red boxes), electroweak data (black line), and recent direct detection analysis of XENON100 after correction of their efficiency factor \cite{XENON100}
}}
\label{fig:relic}
\end{center}
\end{figure}

We show in Fig.\ref{fig:relic} the parameter space still allowed 
after applying all the constraints described above. The red points respect
WMAP constraints after a scan on $m_{\psi_0}$, and the ones below the black
lines are not excluded by electroweak precision tests. The points with black
circle are excluded by the last data released by the XENON experiment.
We observe that a region with $M_{Z_D}\lsim 40$ GeV and 
$10^{-4} \lsim \delta \lsim 10^{-3} $ is still open. We can understand
easily why for increasing values of the kinetic mixing the XENON constraints
seem to weaken: for a fixed $M_{Z_D}$, 
higher values of $\delta$ increase the annihilation cross section, and
decrease the relic density. To fulfill WMAP, one needs to find a point
with $m_{\psi_0}$ far away from the pole ($M_{Z_D}/2$), 
and therefore lighter. This is a region
that XENON has difficulties to exclude:
 the sensitivity of a direct detection experiments decreases
for light dark matter candidate as their efficiencies are worst 
for low-energy nuclear recoil. 
For instance, for $M_{Z_D}=20.6$ GeV and $\delta_1=10^{-4}$, WMAP is fulfilled
for $m_{\psi_0}=10.5$ GeV, which is a point lying exactly in the $Z_D-$pole
region. The spin independent elastic scattering 
on the proton 
is in this case $\sigma^p_{SI} = 7 \times 10^{-40} ~ \mrm{cm^2}$ which is
already excluded by XENON and CDMS-Si. However, for 
$\delta_2=4 \times 10^{-3}$, WMAP is fulfilled for $m_{\psi_0}=4.04$ GeV,
quite avay from the $Z_D$ pole, generating a higher cross section
$\sigma^p_{SI} =  10^{-38} ~ \mrm{cm^2}$ ($\delta_2>\delta_1$)
but which is not yet excluded 
by XENON whose sensitivity is $3.5 \times 10^{-38} ~ \mrm{cm^2}$ for such
a light $\psi_0$.

\subsection{Signals from COGENT, CRESST or DAMA?}
\label{section:Signals}

The DAMA collaboration has provided strong evidence for an annually modulated
signal in the scintillation light from sodium iodine detectors. 
The combined data from DAMA/NaI \cite{DAMA1} (7 annual cycles)
and DAMA/LIBRA \cite{DAMA2} (4 annual cycles) with a total exposure of
0.82 ton yrs shows a modulation signal with 8.2$\sigma$ significance.
The phase of this modulation agrees with the assumption that the signal is due
to the scattering of a WIMP.

Recently, the CoGeNT collaboration has announced
the observation of an excess of low energy events relative to expected background
\cite{COGENT}. This excess, if interpreted as dark matter, implies the dark matter
particles possess a mass in the range of 5-15 GeV and an elastic scattering cross
section with nucleons of the order of $10^{-4}$ pb.
Moreover, recently, a series of analysis and comments
 have been released concerning
the effective value of the XENON100 efficiency  at low energy
($L_{eff}$). We will not enter into all the technical details here, 
a complete analysis  of the computation of $L_{eff}$ and its
consequence on the constraints that we can derive from the XENON experiment
can be found in \cite{X1,X2,X3,X4,X5}.
The main conclusion (until now) is that it is not yet clear if the DAMA/LIBRA and CoGeNT  regions are excluded by XENON100 (see \cite{Fitzpatrick:2010em} for a model independent
analysis concerning light dark matter scenario). 
The main discussion concerns the extrapolation of $L_{eff}$
 and its interpretation in the detection of $S1$ light from low-energy nuclear 
 recoil. To be as conservative as possible, we explore in this section the
 possibility to interpret these excesses with a dark sector with a kinetic
 mixing portal.

\begin{figure}
    \begin{center}
    \includegraphics[width=2.7in]{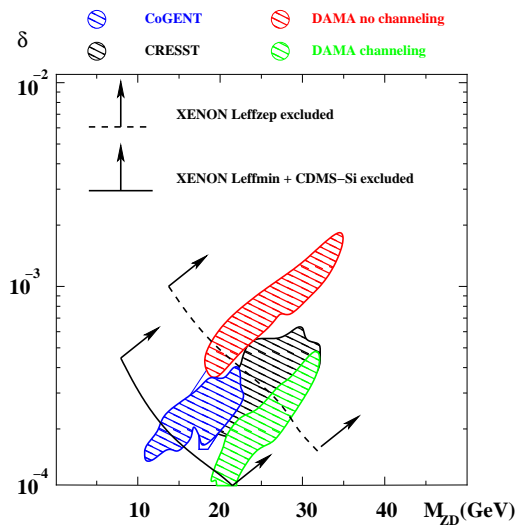}
          \caption{{\footnotesize
Parameter space allowed within 90 \% of  C.L. for the CoGeNT
signal (blue), DAMA without channeling (red), with channeling 
(green), CRESST (black), and the exclusion region depending on the hypothesis
concerning $L_{eff}$ (se the text for details).
}}
\label{fig:Signal}
\end{center}
\end{figure}

 We show in Fig.\ref{fig:Signal} the points respecting 
 WMAP, and the
 DAMA/LIBRA (with and without channeling) CoGeNT and 
 CRESST\footnote{For the CRESST estimation, we used an extrapolation 
 given in the talk of  T. Schwetz and the CRESST 
 collaboration \cite{CRESST}.} results at 90 \% of CL.
 In performing our fits, we have used the 13 DAMA/LIBRA bins below 8.5 keVee
 and the 28 CoGeNT bins between 0.4 and 1.8 keVee. The data at higher
 energies will not include any events from dark matter particles in the mass range
 considered here, and the inclusion of higher energy bins would not affect
 our results in any significant way.
 Concerning the CRESST result, it is important to emphasize that some fraction of the 
 events observed in the oxygen band could be spillage from CRESST's alpha or
 tungsten bands, neutron backgrounds, or be the result of radioactive backgrounds.
 Further information from the CRESST collaboration will be essential for understanding
 these results.
 All the constraints have been calculated for a standard Maxwellian velocity distribution
(with mean velocity 
$v_0=230$ km/s  and an escape velocity $v_{esc}=600$ km/s). 
 One can observe in Fig.\ref{fig:Signal}  that for all experiments, the regions are quite surprisingly
 near and correspond to $10~\mrm{GeV} \lsim M_{Z_D} \lsim 30$ GeV
 and $10^{-4} \lsim \delta \lsim 10^{-3}$, which is in complete agreement
 with the measurement of electroweak precision tests. Moreover, such values
 of $\delta$ are typical of one loop-order corrections and can easily be
 generated by heavy-fermions loops in the $Z-Z_D$ propagator.

\begin{figure}
    \begin{center}
    \includegraphics[width=3.in]{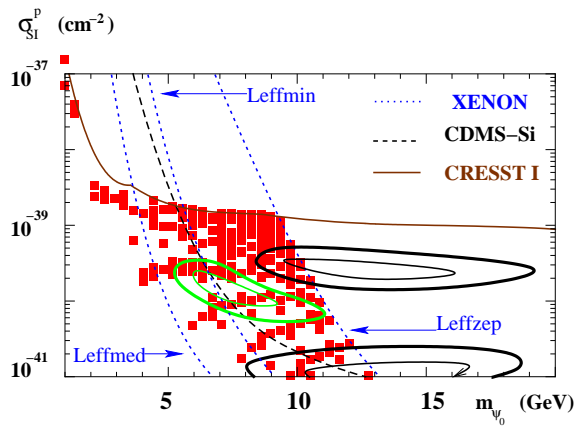}
          \caption{{\footnotesize
Points still allowed by electroweak, cosmological and direct detection
 constraints in the plane ($M_{\psi_0}$; $\sigma^p_{SI}$). The green region corresponds to
  CoGeNT (minimum $\chi^2$, with contours at 90 and 99.9\% C.L.), assuming a constant 
  background contamination \cite{COGENT}. 
  The DAMA regions (goodness-of-fit, also at 90 and 99.9 \% C.L.) are given both with (upper black region) and without (lower black region) channelling 
  \cite{DAMA2}. The black 
  dashed line is the 90 \% C.L. exclusion limit for the CDMS-Si  \cite{CDMS2} 
  and the brown full line the 90\%
  C.L. exclusion limit for the CRESST I experiment \cite{CRESSTI}. The blue dotted lines corresponds to
  the 90\% C.L. exclusion limit from the XENON100 experiment corresponding respectively to
  the LeffMed (left), Leffmin (middle) and LeffZep (right) scintillation efficiency -see text for details. 
}}
\label{fig:Prospect}
\end{center}
\end{figure}

We show in Fig.\ref{fig:Prospect} the points respecting the accelerator,
cosmological, and the more severe direct detection constraints
in the plane ($m_{\psi_0};\sigma^p_{SI}$) in
comparison with XENON100 and CDMS-Si sensitivity.  
To take into account the uncertainties on $L_{eff}$, we plotted 3 exclusion 
limit for XENON corresponding to the best fit set by XENON100 in \cite{XENON100},
which give $L_{eff} \simeq 0.12$ ($L_{effMed}$) at small nuclear recoil energy $E_{nr}$.
A more conservative choice ($L_{effMin}$, corresponding to a lower 90 \% C.L. fit to the data)
gives a $L_{eff}$ which decreases monotonically with $E_{nr}$ and vanishes at 
$E_{nr}< 1$ keV. The ZEPLIN experiment (also a Xe experiment) uses a different $L_{eff}$,
which is essentially zero below 6-7 keV ($L_{effZep}$ here).
For dark matter masses below $\sim 10$ GeV, the CDMS-II silicon detectors provide
very stringent constraints \cite{CDMSII} du to the favorable kinematics of the lighter
target nucleus. However, the observed CDMS-II silicon nuclear recoil quenching
is not reproduced by Linhard theory  \cite{dougherty}. This discrepancy could
indicate a $\sim 20-30$ \% error in the low energy calibration \cite{Hooperlast}. The uncorrected
exclusion curve is also presented in Fig.\ref{fig:Prospect}. 
We also took into account the exclusion limit at 90\% of C.L. from CRESST-I experiment
with sapphire-based cryogenic detector at a threshold of 600 eV \cite{CRESSTI}.
We see that a large region is still to be explored. 
It corresponds to
dark matter masses between 1 and 10 GeV, a range of masses which could be 
difficult but far from impossible to probe in a near future experiment.

\noindent
During the completion of this work, the authors of \cite{Hooperlast} showed that even without channeling but when taking into account uncertainties in the relevant quenching factors, a dark
matter candidate with a mass of approximately $\sim 7$ GeV and a cross section with
nucleons of $\sigma_p^{\mathrm{SI}} \sim 2 \times 10^{-40}$ $\mathrm{cm^{2}}$ could account 
for both these observations. Even if the values they used for the Na quenching factor 
can be considered extreme to some extent, 
these results correspond to 
a dark gauge boson mass $M_{Z_D}\sim 15$ GeV and $\delta \sim 2 \times 10^{-4}$, which is
in the range  of interest for our present study. 
Other interesting constraints to check would be the antimatter production/detection
as computed in specific final states cases in \cite{Lavalle}, galactic gamma-ray 
or from the isotropic diffuse emission \cite{Tytgatbis} and colliders perspectives \cite{Shepherd}.

\section{Conclusion}

We showed that the existence of a $dark$
$U(1)_D$ gauge sector which interacts with the Standard Model only through 
its kinetic mixing possesses a valid dark matter candidate respecting
accelerator, cosmological and the more recent direct detection constraints.
Moreover, considering the latest results of DAMA/LIBRA, CoGENT and 
CRESST, we demonstrated that a specific range of the kinetic mixing 
($\delta \sim  10^{-4}-10^{-3}$) 
can explain all these excesses for a dark boson mass
$M_{Z_D} \sim 10-20 $ GeV.
 Such a value of kinetic mixing is intriguingly 
in agreement with the value one can expect if the mixing is generated by heavy 
hybrid-fermions loop corrections. Other models have similar specificities
(\cite{Tytgat} for instance) : the
diagram for annihilation is the same than the one leading the scattering 
process ($Z_D$ exchange in the former case,
$h$ exchange in the latter).
We also showed that the region of the parameter space still allowed 
by all constraints will be difficult but far from impossible to probe 
in a near future.


\section*{Acknowledgements}
Y.M. wants to thank particularly E. Dudas, T. Schwetz, G. Belanger, N.
Fornengo and A. Romagnoni for useful discussions. The work was
supported by the french ANR TAPDMS {\bf ANR-09-JCJC-0146} 
and the spanish MICINNÕs Consolider-Ingenio 2010 Programme 
under grant  Multi- Dark {\bf CSD2009-00064} and
 the E.C. Research Training Networks under contract {\bf MRTN-CT-2006-035505}.


\end{document}